\documentclass[aps,prl,twocolumn,groupedaddress,showpacs,draft]{revtex4}

\usepackage{mathrsfs}
\usepackage{amssymb}
\usepackage{comment}

\usepackage[intlimits]{amsmath}

\begin{document}

\title{A resolution of the cosmological constant problem}

\author{Jan M. Greben}
\address{CSIR, POBox 395, Pretoria 0001, South Africa; jgreben@csir.co.za, jmgreben@gmail.com}

\begin{abstract}
The standard calculation of vacuum energy or zero point energy is in strong disagreement with observation. We suggest that this discrepancy is caused by the incomplete quantization of standard field theory. The vacuum energy calculation for fermions shows an unacceptable asymmetry between particles and anti-particles, which has to be corrected by an additional quantization step that reverses the order of the anti-particle operators. Boson fields can be handled similarly, but have to be expanded first in terms of bilinear fermion operators. After the full quantization the vacuum energy vanishes. This does not violate the Casimir effect as this can be explained without reference to the vacuum energy, as Jaffe has demonstrated elsewhere.

\end{abstract}



\include {wick.sty}

\maketitle

Quantum field theory (QFT) can boast impressive quantitative successes in its description of nature (see e.g. \cite {Wilczek}). However, the enormous discrepancy between the theoretical calculation of the vacuum energy and observation, known as the cosmological constant problem, remains a serious blemish on its record.
Dirac (quoted in \cite {Susskind}, p. 65) felt that - since every time physicists calculated the vacuum energy it came out infinite - the mathematics must be wrong and that the right answer is that there is no vacuum energy.
Nonetheless, most current physicists seem to have accepted the reality of zero point energy and quantum vacuum energy. Both Weinberg \cite{WeinbRMP} and Caroll \cite{CarrollLRR} quote the Casimir effect \cite{Casimir} as a proof of the reality of zero-point energies. However, Jaffe \cite{Jaffe} has shown that the Casimir effect can just as well be explained by ordinary Feynman diagrams and that the derivation of this effect from the zero point energy - although much simpler than the field theory calculation - is in fact heuristic and hides the dependence on the coupling constant. Jaffe also notes that the question - whether zero-point fluctuations of quantized fields are real - has a long history, with some prominent physicists arguing against its existence.
In this paper we argue that the dominant vacuum energy term in QFT which is responsible for the cosmological constant problem is an artifact of the quantum calculation and due to a missing foundational step in QFT. The need for new QFT foundations to deal with this discrepancy has been suggested before (e.g. by \cite {Abbott}), but such a step should not endanger the impressive quantitative successes of QFT already obtained. For the cases considered our new formulation satisfies this criterion, which shows once again that there are often different routes in physics to reach the same quantitative successes.

We start with a discussion of the fermionic case, where there is clear evidence that there is something wrong with the standard calculation of the vacuum energy. The quantized fermion field $\psi(x)$ is typically written as follows:
(\cite {Itzykson}, p.145; \cite{Bogoliubov83}, p.85; \cite{Peshkin}, p.54; \cite{Schweber}, p.228):
\begin{eqnarray}
  \label{eq:expansion}
\psi(x)=\frac{1}{(2\pi)^{3/2}}\sum_\alpha\int
d^3\vec{q} \left(\frac{m_\alpha}{E_{\vec{q}}}\right)^{1/2}
 \nonumber\\
 \times \left[b_{\alpha\vec{q}}\,
u_{\alpha\vec{q}}\,e^{-iqx}+d^\dag_{\alpha\vec{q}}\,v_{\alpha\vec{q}}\,e^{iqx}\right],
\end{eqnarray}
where we used Sakurai's \cite{Sakurai} convention to indicate particle operators with the symbol $b_{\alpha\vec{q}}$  and anti-particles operators with $d_{\alpha\vec{q}}$.
The creation and annihilation operators satisfy the anti-commutation rules:
\begin{equation}
  \label{eq:expansion}
\{b_{\alpha\vec{p}},b^\dag_{\beta\vec{q}}\}=\delta_{\alpha\beta}\delta(\vec{p}-\vec{q}),~~~~~~~
\{d_{\alpha\vec{p}},d^\dag_{\beta\vec{q}}\}=\delta_{\alpha\beta}\delta(\vec{p}-\vec{q}),
\end{equation}
with all other anti-commutators zero.
The index $\alpha(\beta)$ represents the usual discrete fermion
quantum numbers, e.g. spin, flavor and - in the case of quarks - color.
The fermionic vacuum energy - i.e. the spatial integral over $T_{00}(x)$ - can be expressed as an integral over the fermion field and reads (e.g. \cite{Sakurai}, p.150):
\begin{eqnarray}
 \label{eq:vacuumfermion}
  E_0=\sum_\alpha \int d^3q\, q_0 < 0| b^\dag_{\alpha\vec{q}} b_{\alpha\vec{q}}\,
 -\,d_{\alpha\vec{q}} d^\dag_{\alpha\vec{q}}|0 >=
 \nonumber\\=
 \sum_\alpha \int d^3q\, q_0 [<0| b^\dag_{\alpha\vec{q}} b_{\alpha\vec{q}}\
 +\,d^\dag_{\alpha\vec{q}} d_{\alpha\vec{q}}|0 >-\frac{V}{(2\pi)^3}]
\end{eqnarray}
where we replaced $\delta^{(3)}(0)$ by $V/(2\pi)^3$. The last term is the fermionic vacuum energy. It is either infinite or enormous if we use the Planck energy as cut-off. The important thing to notice is that
 only anti-particles contribute to this vacuum energy. This enormous asymmetry between particles and anti-particles in this elementary quantum calculation is not acceptable.
  Since the particle term gives the natural zero result for the vacuum energy, one must conclude that the quantum treatment of the anti-particles is deficient. However, the procedure to combine anti-particle creation operators with particle annihilation operators in the field $\psi$, and the conjugate combination in $\bar{\psi}$, is well-founded and leaves no room for change. We thus must conclude that the usual process of transcribing the classical fields into quantum operators is not sufficient to capture the full quantization process when both particles and anti-particles are present.

 To analyze this problem in more detail we consider a one-particle matrix element $<\beta|b^\dag_\delta b_\gamma|\alpha>$ in the expansion of a typical field expression $\bar{\psi} O\psi$. As expected this matrix element is associated with the one particle matrix element $\bar{u}_{\beta} O \,u_{\alpha}$. If we do the same for anti-particles, we must consider $-<\beta| d_\gamma d^\dag_\delta |\alpha>$, with $\alpha$ and $\beta$ now anti-particle states. After (anti-)commuting the operators we get the natural result $\bar{v}_{\alpha} O \,v_{\beta}$, plus an extra term without clear physical significance. The natural term is similar to the particle result, except that the "initial" and "final" state have been interchanged. This was to be expected, as anti-particles behave in some way as particles moving backwards in time. However, time ordering does not exist in the current case with a single space-time variable, and initial and final now refer to the ket and bra state vector, with the ket vector seen as the initial state for particles. In keeping with this generalized notion of the ordering of anti-particle operators for a single space-time point, we suggest that the proper treatment of the operator expressions between state vectors is to reverse the order of the anti-particle operators taking into account the anti-commutativity of the operators, but ignoring the anti-commutators. In this way the extra (unphysical) term(s) will not appear and the expression takes on a fully symmetric form in the order natural to anti-particles. This reversal will be called the $\mathbb{R}$-product and was already discovered before in the context of self-consistent field equation solutions \cite{Greben}. We stress that the $\mathbb{R}$-product applies only to operators at the same space-time point. There would be no justification for adjusting the order of the anti-particle operators belonging to different space-time coordinates, as the Wick expansion already controls the order in that case. For the anti-particle contribution to the vacuum energy we now get:
\begin{eqnarray}
 \label{eq:Rproduct1}
 < 0| \mathbb{R} \{-\,d_{\alpha\vec{q}} d^\dag_{\alpha\vec{q}}\}|0 >=
   < 0| d^\dag_{\alpha\vec{q}} d_{\alpha\vec{q}}|0 >  ~=0\,.
\end{eqnarray}
Hence after restoring the symmetry between particles and anti-particles, the vacuum energy automatically vanishes. This is the natural result, as already expected by Dirac. Because this $\mathbb{R}$-product must be applied generally, it will also eliminate the charge and momentum vacuum terms.

The usual way to eliminate the vacuum terms is via the normal product (\cite {Kaku}, \cite{Sakurai}, \cite{Schweber}). However, this product is seldom seen as a principled procedure. One simply argues (e.g. \cite{Peshkin}) that the absolute energy level plays no role in QFT, so that one is justified in omitting the vacuum term by applying the normal product. Some authors go further (e.g. \cite {Bogoliubov59}, p. 103-104; \cite{Lahiri}, p. 82), stating that the original Lagrangian must be normal ordered before entered into the Wick expansion. However, they also fall short in justifying this step physically. Because the obscure physical justification of this product (as it is characterized by \cite {Jaffe}) one re-admits the vacuum energy in cosmological considerations, ignoring the normal product and thus re-introducing the vacuum energy leading to the enormous discrepancy. Instead, the $\mathbb{R}$-product is a mandatory quantization step.
For simple operator expressions like Eq. (\ref{eq:Rproduct1}), the two products yield identical results, so that the $\mathbb{R}$-product in those cases justifies the application of the normal product in this context.
For more complex expressions the two products yield very different results, as we will see shortly.

We have presented the fermionic case before the boson case as the inadequacy in the fermionic case is more obvious. In standard QFT text books the boson case is introduced first as it is considered simpler and more instructive. However, we will argue that this simplicity is misleading and that a more subtle treatments of the boson case is required.
Schematically, the standard Hamiltonian for the boson or harmonic oscillator case reads:
 \begin{eqnarray}
  \label{eq:aa}
  H=\frac{1}{2}\sum \hbar\omega(a^\dag_{\vec{k}}\, a_{\vec{k}}+a_{\vec{k}} \;a^\dag_{\vec{k}}\ )=
  \nonumber\\=
  \sum \hbar\omega \,a^\dag_{\vec{k}}\, a_{\vec{k}}+\frac{1}{2}\frac{V}{(2\pi)^3} \sum \hbar\omega,
 \end{eqnarray}
where the last term represents the vacuum energy. This term diverges like $\Lambda^4$, where $\Lambda$ is the cut-off in momentum space. The usual argument \cite{CarrollLRR} is that the only reasonable cut-off is of the order of the Planck energy, making this vacuum energy density enormous. In contrast to the fermionic case there seems nothing obviously wrong with this result. As usual one can remove the vacuum term with the normal product, but since this product has no physical legitimacy this procedure does not resolve the cosmological constant problem. However, there is a more subtle way to effect the quantization in the boson case, leading to the elimination of the vacuum energy term using the $\mathbb{R}$-product.

Although historically bosons (like the photon) play a very fundamental role in physics, in modern gauge theory the
boson fields are postulated and required to realize certain gauge symmetries of the fermionic Lagrangian. As such these boson fields are secondary, as they are only needed because of the presence of fermion fields. The equations of motion confirm this: the source term of the boson fields has the typical structure $\bar{\psi} O \psi$, which suggests that the basic operator structure of boson fields can be expressed as a bilinear expansion in fermion fields. This is exactly, what was found in the self-consistent solution of quantum field equations \cite{Greben}. Similar fermion expansions are commonplace in nuclear structure physics, where bosonic excitations are expanded in terms of particle-hole configurations (see e.g. \cite {Shalit}), although there are essential differences between the two cases.
Clearly, the underlying fermions in such an expansion should be massless to enable the construction of massless gluon and photon fields (QCD and QED). Hence, these fermions cannot be identified with the massive fermions of the standard model and must have a more fundamental massless status. Such basic massless quarks also underly the self-consistent calculations mentioned above \cite{Greben}, where the massive standard model quarks only emerge after dressing the bare quarks through the interactions with other fields.

The explicit bilinear expansion will be discussed in more detail later, but let us first consider the consequences of such an expansion for the boson vacuum energy schematically. We write symbolically:
\begin{equation}
  \label{eq:R5}
  a^\dag_{\vec{p}}\, \rightarrow\, b_{\alpha}^\dag d_{\beta}^\dag~~\textrm{and}~~
 a_{\vec{p}}\,\rightarrow\,d_{\beta'} b_{\alpha'}~.
 \end{equation}
Now let us rewrite Eq. (\ref{eq:aa}) using this symbolic notation. Since all
operators in Eq. (\ref{eq:aa}) refer to fields at the same space-time point we must
apply the $\mathbb{R}$-product. We obtain:
\begin{eqnarray}
  \label{eq:R7}
  \case{1}{2} \int d\vec{p}~p_0
  \left\{a^{\dag} _{\vec{p}}\,a_{\vec{p}}\,+a_{\vec{p}}\,a^\dag _{\vec{p}}\right\}\rightarrow
    \nonumber \\
      \mathbb{R}[b_{\alpha}^\dag d_{\beta}^\dag \,d_{\beta'} b_{\alpha'}+
  d_{\beta'} b_{\alpha'} b_{\alpha}^\dag d_{\beta}^\dag]\,=
  \nonumber \\
=     -b_{\alpha}^\dag d_{\beta'} d_{\beta}^\dag \, b_{\alpha'}
    -d_{\beta}^\dag \,b_{\alpha'}\,b_{\alpha}^\dag\, d_{\beta'}\,.
 \end{eqnarray}
 The vacuum matrix element of the right-hand side is
 zero: in the first term because of the presence of $b_{\alpha'} $ on
 the right; in the second term because of the operator
 $d_{\beta'}$ on the right. Hence, the boson vacuum energy vanishes under this boson representation. The final expression displays a beautiful symmetry between particles and anti-particles, represented by the symmetry $b \leftrightarrow d$.
 In this case the normal product would lead to:
 \begin{equation}
  \label{eq:normal}
  :b_{\alpha}^\dag d_{\beta}^\dag \,d_{\beta'} b_{\alpha'}+
  d_{\beta'} b_{\alpha'} b_{\alpha}^\dag d_{\beta}^\dag:=     2 b_{\alpha}^\dag d_{\beta}^\dag \,d_{\beta'} b_{\alpha'},
 \end{equation}
  which does not show the same elegant symmetry. Under the normal product the two initial terms loose their independent character and are reduced to the same expression. The same phenomenon holds for longer expressions: the $\mathbb{R}$-product maintains the individual character of each term and ensures the symmetry between terms, while the normal product does not necessarily maintain the distinction between different terms. Hence, in addition to having a strong physical justification, the $\mathbb{R}$-product also yields a much more elegant formulation than the normal product. 
  
  These results indicate that the boson operator expansion in terms of boson operators $a$ and $a^\dag$ is incapable of fully capturing the quantization process.
  The required fermionic expansion suggests that the boson fields are composite fields, however this is not true in the traditional sense, as the composing fermionic fields are defined at the same space-time point and with the same momentum (see upcoming Eq. (\ref{eq:AJMG})). In most respects the boson fields maintain their fundamental status, for example in the functional derivation of the field equations. Nonetheless, giving up some degree of fundamentality for the boson fields is already a considerable step from a historical perspective and illustrates the depth to which one has to go to resolve the cosmological constant problem.

 An important question raised by this proposal is whether the new fermionic representation of bosons is able to reproduce the same magnificent quantitative QFT results as the standard boson representation. A related question is whether the new representation of bosons provides new insights in those problems in QFT which have not (yet) been explained satisfactory or could only be handled through rather artificial technical manipulations, such as the Gupta-Bleuler formulation \cite{Gupta}, \cite{Bleuler}. A full discussion of these questions lies outside the realm of this short paper. It is not unusual that the same physical phenomenon can be represented by different models or theories with the differences or inadequacies only showing up under special circumstances. We already saw above that the Casimir effect can be explained in different ways. Initial indications are that some standard results (such as the photon propagator) are easily derived within this new representation.
 
 To illustrate this new representation let us present the suggested form of the electromagnetic field. The usual representation is (\cite{Itzykson}, p. 129; \cite{Peshkin}, p. 123; \cite{Sakurai}, p. 132):
  \begin{eqnarray} \label{eq:Astandard}
A_\mu(x) \propto \int \frac{d^3p}{E^{1/2}}\sum_{\lambda=0}^3
\left\{a^{(\lambda)\dag}(\vec{p})\,e^{(\lambda) *}_{\mu}(\vec{p}) e^{ipx}+c.c.\right\},
 \end{eqnarray}
 where $\lambda=1$ and $2$ correspond to the physical transverse waves, whereas  $\lambda=0$ and $3$ correspond to the scalar and longitudinal component.
 We now must replace this with a similar functional expression in terms of fermionic creation and annihilation operators:
\begin{eqnarray} \label{eq:AJMG}
A_\mu(x) \propto \sum_{\alpha,\beta}\int \frac{d^3p}{E^{1/2}}
\left\{ (\bar{u}_{\alpha,\frac{\vec{p}}{2}} \gamma_{\mu} v_{\beta,\frac{\vec{p}}{2}})b^\dag_{\alpha,\frac{\vec{p}}{2}}\,d^\dag_{\beta,\frac{\vec{p}}{2}}e^{ipx}+c.c.\right\}.
\end{eqnarray}
The free fermionic spinors $u$ and $v$ are initially defined for finite $m$, but taken to the limit $m\rightarrow 0$ in the end. This procedure ensures that the amplitude has the correct final structure.
The Lorentz vector structure is now automatically implemented. The Lorentz condition $\partial^\mu A_\mu(x)=0$ is also automatically satisfied as $p^\mu(\bar{u}_{\alpha,\frac{\vec{p}}{2}} \gamma_{\mu} v_{\beta,\frac{\vec{p}}{2}})=0$.
One can show that this amplitude yields the correct electromagnetic propagator in the limit that $m\rightarrow 0$. The derivation of the propagator is not affected by the $\mathbb{R}$-product as the space-time coordinates in the vacuum expectation value (vev) defining the propagator are not identical. For QCD we can generalize this expression by including the operator $\lambda_a$ in the matrix elements. However, for QCD and other non-linear theories the bilinear expansion in Eq. (\ref{eq:AJMG}) no longer suffices as the fermionic operators are iterated to higher order through the solution of the quantized field equations. A complete expansion has been carried out for the dressing of single quarks in Ref. \cite{Greben} in a bound-state setting. A similar expansion should hold for scattering amplitudes.

In principle the fermionic expansion allows for the occurrence of a new type of bosonic quantum amplitude which has no explicit space-time dependence. An example is given below:
\begin{eqnarray} \label{eq:AConsJMG}
A(x) \propto \sum_{\alpha,\beta}\int d^3p
 \left\{ (\bar{u}_{\alpha,\frac{\vec{p}}{2}} O u_{\beta,\frac{\vec{p}}{2}})b^\dag_{\alpha,\frac{\vec{p}}{2}}\,b_{\beta,\frac{\vec{p}}{2}}\right.
 \nonumber \\
+\left.(\bar{v}_{\alpha,\frac{\vec{p}}{2}} O v_{\beta,\frac{\vec{p}}{2}})d_{\alpha,\frac{\vec{p}}{2}}\,d^\dag_{\beta,\frac{\vec{p}}{2}}\right\},
\end{eqnarray}
where we did not specify the structure of the operator $O$. For a scalar operator $O$ this amplitude behaves very much like a constant times an identity operator, so that this operator embedded within a string of other operators acts effectively like a constant. Such terms could play an important role in the satisfaction of the quantum field equations for non-linear fields.
A quantum operator of this nature could well assume the role that the constant Higgs 'vev' plays in standard theories.
A more dynamic origin of the Higgs mechanism is necessary as the Higgs 'vev' itself will be zero in the current framework, ensuring the vanishing of a Higgs contribution to the vacuum energy, as well.
These constant amplitudes would not have any direct influence on the scattering processes.

It is clear that many technical issues remain due to the required boson expansion. However, it would be somewhat naive to expect that the biggest discrepancy in physics ever recorded could be resolved without having some deep consequences for QFT.
The presence of vacuum energy is still possible as a consequence of a non-zero cosmological constant. Cosmological considerations indicate that the value of the vacuum energy density is about $\epsilon=4.1\times 10^{-47}\,\mathrm{GeV}^4$ \cite {GrebenCos}, corresponding to a cosmological constant of $\Lambda=8\pi G\epsilon= 6.7\times 10^{-84}\,\mathrm{GeV}^2$. The relationship of these constants to particle physics are governed by the ratio
$(\epsilon/G)^{1/6}= 43.1 \,\mathrm{MeV}$. This ratio appears naturally when the self-consistent QFT treatment of single quarks is combined perturbatively with general relativity \cite {Greben} and leads to a realistic estimate of the light quark masses of the standard model.

\end{document}